\documentclass[letterpaper,twocolumn,10pt]{article}
\usepackage{usenix2019_v3}

\usepackage{tikz}
\usepackage{amsmath}
\usepackage{graphicx}
\usepackage{booktabs}
\usepackage{multirow}
\usepackage{array}
\usepackage{adjustbox}

\usepackage[linesnumbered,ruled,vlined]{algorithm2e}
\usepackage{mathtools}
\usepackage{graphicx}
\usepackage{booktabs}
\usepackage{multirow}
\usepackage{array}
\usepackage{adjustbox}
\usepackage{threeparttable}
\usepackage[normalem]{ulem}
\useunder{\uline}{\ul}{}
\SetKwInput{KwInput}{Input}
\SetKwInput{KwOutput}{Output}

\graphicspath{{Images/}}

\begin{document}

\date{}

\title{\Large \bf Graph-based Solutions with Residuals for Intrusion Detection:\\
  the Modified E-GraphSAGE and E-ResGAT Algorithms}


\author{
{\rm Liyan Chang}\\
The University of Hong Kong
\and
{\rm Paula Branco}\\
University of Ottawa
} 

\maketitle

\begin{abstract}
The high volume of increasingly sophisticated cyber threats is drawing growing attention to cybersecurity, where many challenges remain unresolved. Namely, for intrusion detection, new algorithms that are more robust, effective, and able to use more information are needed. Moreover, the intrusion detection task faces a serious challenge associated with the extreme class imbalance between normal and malicious traffics. Recently, graph-neural network (GNN) achieved state-of-the-art performance to model the network topology in cybersecurity tasks. However, only a few works exist using GNNs to tackle the intrusion detection problem. Besides, other promising avenues such as applying the attention mechanism are still under-explored.

This paper presents two novel graph-based solutions for intrusion detection, the modified E-GraphSAGE and E-ResGAT algorithms, which rely on the established GraphSAGE and graph attention network (GAT), respectively. The key idea is to integrate residual learning into the GNN leveraging the available graph information. Residual connections are added as a strategy to deal with the high class imbalance, aiming at retaining the original information and improving the minority classes' performance. An extensive experimental evaluation on four recent intrusion detection datasets shows the excellent performance of our approaches, especially when predicting minority classes.

\end{abstract}

\section{Introduction}
Cybersecurity has gained increasing attention in the contemporary society. Due to the easy access to the global network, individuals and organizations are faced with more complex cyber threats that arrive with at an increasing pace. The frequency and the intricacy of network attacks have significantly risen over the past few years posing a serious challenge. 

Intrusion detection system (IDS), a model that identifies potentially malicious network traffic, plays an important role in cybersecurity~\cite{liao2013intrusion}. Machine learning models are increasingly used in this environment and the recent upsurge of deep learning techniques resulted in major advancements for IDSs. Neural networks are indeed playing a central role showing vast progress concerning IDS development (e.g. \cite{aldwairi2018evaluation,ullah2019two,yin2017deep}). In particular, graph neural networks (GNNs) which only recently started to be explored in cybersecurity tasks, are achieving state-of-the-art performance in many cases~\cite{jiang2021graph}. Yet, the number of solutions for intrusion detection problems that rely on GNNs is still scarce. Although this is a promising avenue, it is still under-explored. Moreover, a critical issue remains to be solved: the serious class imbalance present in the majority of real-world cybersecurity problems. Most of the current IDSs have suffered from the extreme imbalance between the normal and malicious traffic, and a good effective solution is still missing.

GNNs are designed to encode non-Euclidean graph data, such as social networks, biological networks, and citation networks, to facilitate representation learning and downstream classification and prediction tasks. A graph-structured dataset normally contains node and/or edge features. An effective GNN not only learns feature embeddings, but it also captures the spatial information hidden in the graph topology. Intrusion detection problems are typically performed on network flow data, which is comprised of IP address, port numbers and a set of flow-related features (the flow duration, transaction bytes, and the number of transmitted packets, etc). GNNs provide a perfect application context for intrusion detection problems by simply mapping IPs and ports to nodes while network flows can be mapped to edges. 

In this paper, we extend the idea of E-GraphSAGE~\cite{lo2021graphsage} to explore more alternatives of GNNs, such as graph attention networks (GATs), and propose two GNN solutions for intrusion detection tasks that leverage the residual learning. The first proposal builds upon E-GraphSAGE by combining it with residual learning, in order to tackle the high class imbalance presented in the datasets. Inspired by the improvements observed on this solution, we introduce a second novel proposal named edge-based residual graph attention network (E-ResGAT). E-ResGAT uses the attention mechanism of GAT while also supporting edge features and embedding residual learning to further enhance the performance. We demonstrate that these graph-based models with residuals can effectively detect malicious traffic and deal with the class imbalance problem, through an extensive experimental evaluation on four well-known intrusion detection datasets. Our results show that our proposed models achieve better performance than the original models and provide important improvements in the classification of the minority classes. 

The rest of this paper is organized in the following way. Section~\ref{sec:2} describes the related work on previous graph-based models in the cybersecurity tasks. Section~\ref{sec:3} presents the graph construction from the traffic network and provides a theoretical overview of both proposed graph-based algorithms with residuals. Section~\ref{sec:4} provides the experimental evaluation details including a summary of four intrusion detection datasets, the experimental settings, the results obtained and a discussion. Finally, concluding remarks and the future work are presented in 
Section~\ref{sec:5}.

\section{Related Work}\label{sec:2}


Graph-based deep learning is a fast evolving and promising area of research that is witnessing a growth in new solutions and applications. Recently, graph neural networks (GNNs)~\cite{scarselli2008graph} have achieved state-of-the-art performance in cybersecurity tasks, such as network intrusion and anomaly detection~\cite{jiang2021graph}.
They are suitable for cybersecurity tasks, owing to their ability to capture the spatial information hidden in the network topology and can be generalized to unseen topologies when the networks are dynamic. Compared to traditional neural networks, these models take both graph topology and flow features into consideration. Usually, hosts are viewed as nodes, while flows in between are regarded as edges in graph-based models. This section covers a wide range of graph learning techniques used for intrusion detection, including graph embedding, graph convolutional and graph attention networks.

Graph embedding transforms the edges, nodes and their features information into a vector space. Multiple approaches involving graph embedding have been proposed, including Structure2Vec~\cite{dai2016discriminative} and GraphSAGE~\cite{lo2021graphsage}, among others. Structure2Vec~\cite{dai2016discriminative} is based on the idea of embedding latent variable models into feature spaces, and has shown to be efficient and scalable for structured data representation, although it was not applied to the network intrusion detection problem. 

A graph embedding approach that learns the representations of network flows in an anomaly detection task was proposed by Xiao et al.~\cite{xiao2020towards}. Network flows begin by being transformed into two graphs. The first graph learns the embedding by using its own IP address and port number, while the second graph generates the embedding from all IPs and port numbers. This method captures both local and global embedding representations of network flows. However, it uses a transductive method, which learns the representation for each individual flow. Thus, this model may fail to classify unseen network flows. Our proposed models show an advantage as both learn the neighborhood aggregation, and are thus applicable to inductive tasks as well. The GraphSAGE network proposed by Hamilton et al.~\cite{hamilton2017inductive}
is an example of an inductive approach.
Although allowing the generation of low dimensional vector representation of graphs, this algorithm still has a significant drawback as it does not consider edge features. GraphSAGE has been applied to several problems such as network slicing management and traffic prediction. However, its application to network intrusion detection, our target problem, was proposed after the introduction of a modification to the GraphSAGE algorithm that we describe next.

The extension of GraphSAGE to the intrusion detection problem was carried out by Lo et al.~\cite{lo2021graphsage} that proposed E-GraphSAGE algorithm. This is an inductive model using GraphSAGE network that also allows to take into consideration the edge features. The authors first aggregate the neighboring flows for both source and destination nodes and then concatenate them as the flow representation. This model learns the aggregation function rather than the direct flow representation and thus, it can be easily applied to unseen flows. However, one of its disadvantages is that the flow feature itself does not have enough impact on the embedding representation. Our first proposed model addresses this issue in E-GraphSage, by proposing a modification that allows the algorithm to act as residual learning. We focused on E-GraphSAGE to carry out this modification given its suitability for the intrusion detection problem.

Graph convolutional networks (GCN) extend to the graph scenario with the convolution operation that is used on traditional data such as images~\cite{jiang2021graph}. The key idea of GCNs is to learn a function that maps the graph information into a new representation. In this function, a node aggregates not only its own features but also its neighbors features in order to generate the new representation of the data. Cheng et al.~\cite{cheng2021discovering} proposed Alert-GCN, a solution that aims at correlating alerts that belong to the same attack using GCNs. Alert-GCN tackles this task as a node classification problem. It starts by building an alert graph using the alert information from neighbors which is then feed into GCN to perform node classification. Zhou et al.~\cite{zhou2020automating} performed botnet detection using a graph convolutional network, which first generated botnet traffic by creating botnet connections mixed with different real large-scale network traffic flows. Then, they applied the graph convolutional network for network node classification. The generated graph does not include any flow or node features, and only considers the topological information of the network connectivity graph. Furthermore, the approach does not detect individual attack flows, but is limited to detecting attack nodes, specifically in the context of botnets. In contrast, We utilize both topology information and edge features in our proposed models, and aim to detect a variety of attack families.

Graph attention networks~\cite{velivckovic2017graph} (GAT) are widely applied in communication network and several studies have shown their potentials. The key distinguishing aspect of GAT models lies on the incorporation of the attention mechanism into the propagation step while using the multi-head attention mechanism to make the learning process more stable.
One of the advantages of the attention mechanism is that it focuses on the most relevant neighbors of the flow to make decisions, rather than taking the neighbors equally. Multi-scale Spatial-Temporal Graph Neural Network (MSTNN)~\cite{yang2020mstnn} uses the attention mechanism on the spatial extractor for capturing the time-varying spatial correlations of nodes in origin-destination traffic prediction task. Wan et al. proposed GLAD-PAW~\cite{wan2021glad} for anomaly detection in log files. GLAD-PAW contains a position aware weighted graph attention layer, which encodes the node features with position information. However, GAT is seldom applied in intrusion detection and only a few recent research works use it in this context. Our second proposed method includes an attention mechanism and demonstrates good generalization ability on three current intrusion detection datasets.

\section{Our Proposed Graph-based Algorithms with Residuals}\label{sec:3}
This section first introduces the graph construction from the network flow data and then presents our two proposed algorithms: the modified E-GraphSAGE model and the E-ResGAT model. We use the following notation: vectors are represented with bold lowercase letters and matrices are denoted with bold uppercase letters.

\subsection{Graph Construction}\label{sec:3_1}

Network flow data consist of the following three components, source and destination addresses, and flow features, including duration, transaction bytes, transmitted packet sizes, etc. It is natural to construct a graph using source and destination as the endpoints of a network flow. To be more specific, source IP addresses and port numbers are combined to identify graph nodes, and so are the destination addresses and ports. The remaining flow data serves as the edge features. In this way, the intrusion detection problem is encoded as an edge classification task. 

\begin{figure}[htb]
    \centering
    \includegraphics[width=0.45\textwidth]{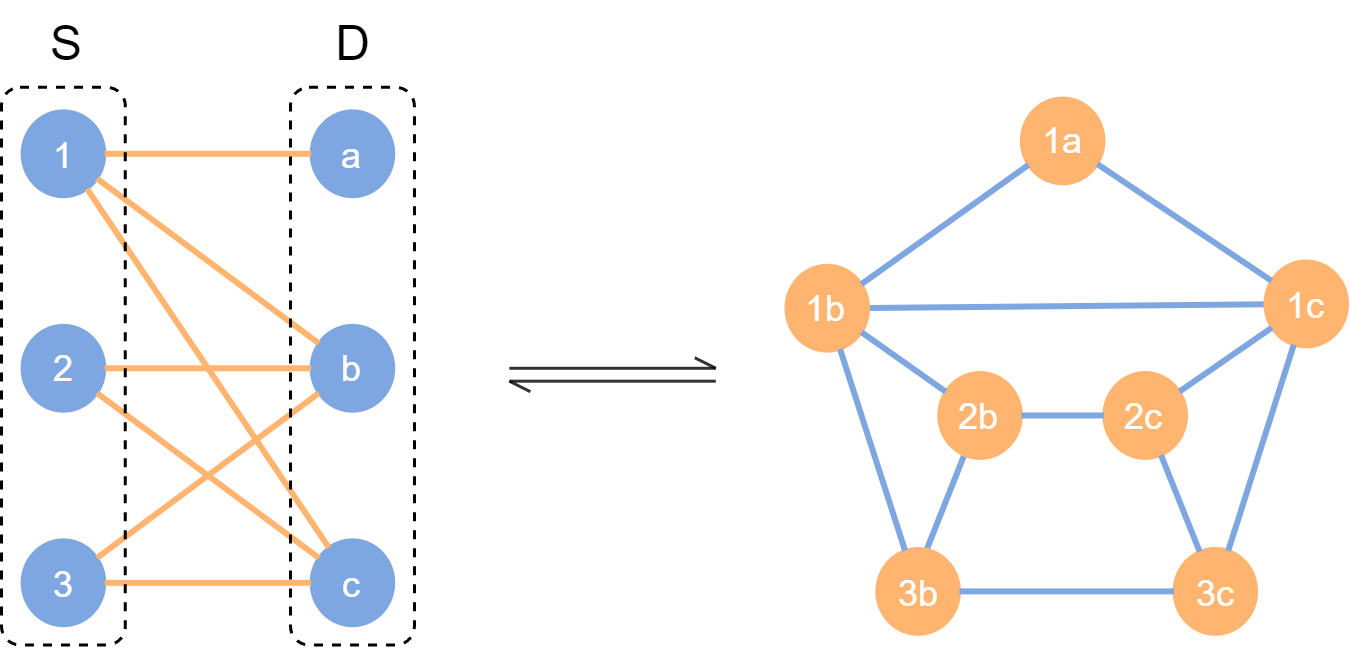}
    \caption{The transformation between the original bipartite graph (left) and the corresponding line graph (right). Nodes on the right correspond to the edges on the left. $\mathcal{S}$ refers to a source node set of size 3, and $\mathcal{D}$ a destination node set of the same size.}
    \label{fig:bigraph}
\end{figure}

Since in all of our datasets, the source and destination addresses are disjoint, we construct a bipartite graph $\mathcal{G(S,D;E)}$, where $\mathcal{S}$, $\mathcal{D}$, $\mathcal{E}$ represent the source node set, destination node set, and edge set, respectively. Notice that the graph nodes are featureless. We can easily convert the bipartite graph into its line graph counterpart~\cite{lehot1974optimal}, whose nodes correspond to the original edges and vice versa. Thus, the problem becomes a node classification task. Figure~\ref{fig:bigraph} shows a simple example of such transformation. Our proposed modified E-GraphSAGE and E-ResGAT models are based on the described bipartite and line graph structures, respectively. 

Furthermore, we augment the source (resp. destination) node set to the size of destination (resp. source) node set, by creating virtual nodes, if $|\mathcal{S}| < |\mathcal{D}|$ (resp. $|\mathcal{S}| > |\mathcal{D}|$). Then, the source (resp. destination) endpoints of the edges are randomly replaced with the virtual nodes. The purpose for the augmentation is two-fold. Firstly, as the full neighborhood is used in the GAT-based models, the adjacency list of the line graph may not fit in the memory. This step deliberately increases the number of nodes and therefore reduces the averaged node degree, since the number of edges $|\mathcal{E}|$ remains unchanged. Note that $|\mathcal{E}|$ in the line graph is given by $\sum_{i\in \mathcal{S}\cup \mathcal{D}}(d_i-1)d_i/2$, where $d_i$ is the degree of node $i$. In this way, the memory of line graph will be reduced, especially when there exists a large node degree. Secondly, the node augmentation introduces a random mapping, which will prevents the potential issue of the source nodes providing an unintentional label for malicious traffics~\cite{lo2021graphsage}.


\subsection{The Modified E-GraphSAGE Algorithm}\label{sec:egrap}

The E-GraphSAGE algorithm~\cite{lo2021graphsage} is adapted from the GraphSAGE algorithm~\cite{hamilton2017inductive} to support edge classification. The key idea behind E-GraphSAGE is that it performs GraphSAGE algorithm on two endpoints of an edge separately, and finally concatenate node embeddings as the edge representation. Our key contribution lies on a modification on the final concatenation step that aims at overcoming the potential loss of information of the original edge features.

Similar to the E-GraphSAGE, our proposed modified E-GraphSAGE algorithm consists of two parts, sampling and aggregation. Due to the memory limitation, we implemented the mini-batch version of both the original and the modified E-GraphSAGE algorithm. The sampling and aggregation components of our modified E-GraphSAGE algorithm are shown in Algorithm~\ref{alg:esamp} and~\ref{alg:eagg}, respectively. 

\paragraph{\textbf{\textup{Sampling}}}
In the modified E-GragpSAGE, we uniformly sample a fixed-size set of neighbors, instead of using full neighborhood as in \cite{lo2021graphsage}. We define the neighborhood $\mathcal{N}_v$ of a node set $\mathcal{V}$, as a fixed-size, uniform sample from the set $\{u : v\in \mathcal{V}, uv \in \mathcal{E}\}$, where $\mathcal{V} \subseteq \mathcal{S} \cup \mathcal{D}$. For each edge batch $\mathcal{B}$, we draw different uniform samples at each iteration $k$ as shown in Equation~\ref{eq:batchUsamp}.

\begin{equation}\label{eq:batchUsamp}
    \mathcal{B}^{k}\leftarrow\mathcal{B}^{k}\cup\{uv, \forall{u} \in\mathcal{N}_v\}
\end{equation}

Without this sampling, the memory and expected runtime of a single batch is unpredictable and in the worst case $O(|\mathcal{V}|)$~\cite{hamilton2017inductive}. In practice, the average degree of nodes is used as the size of neighborhood and so, we sampled a 2-hop 8-neighborhood of the batch before aggregation. 

\begin{algorithm}[ht]
\SetAlgoLined
\KwInput{   Graph $\mathcal{G(V,E)}$\;  
\Indp\Indp edge minibatch $\mathcal{B}$\;
node set $V(\mathcal{B})=\{v:v$ is an end point of $uv \in \mathcal{B}\}$\;
depth $K$; 
neighborhood sampling functions $\mathcal{N}_v:v\rightarrow2^\mathcal{V}$\;
}
\KwOutput{2-hop neighborhood of the batch $\mathcal{B}^0$}
 \BlankLine
 $\mathcal{B}^{K}\leftarrow\mathcal{B}$\;
 \For{$k=K,...,1$}{
  $\mathcal{B}^{k-1}\leftarrow\mathcal{B}^k$\;
  \For{$v \in V(\mathcal{B}^k)$}{
  $\mathcal{B}^{k-1}\leftarrow\mathcal{B}^{k-1}\cup\{uv, \forall{u} \in\mathcal{N}_v\}$\;
  }
 }
 \caption{E-GraphSage minibatch sampling}
 \label{alg:esamp}
\end{algorithm}

\paragraph{\textbf{\textup{Aggregation}}}
After sampling, the algorithm iteratively aggregates the neighboring edge features layer by layer. We now describe a single layer below. The input to the layer are the edge features $\mathbf{h}_{uv}$, where $uv \in \mathcal{B}^0$. At the $k^{th}$ layer, the neighboring edge features in the previous layer are aggregated to the node $v$, which can be expressed as follows:

\begin{equation}\label{eq:neighedge}
    \mathbf{h}_{\mathcal{N}_v}^k = AGG_k(\{\mathbf{h}_{uv}^{k-1},\forall{u}\in \mathcal{N}_v\})
\end{equation}

In Equation~\ref{eq:neighedge}, various kinds of aggregation functions $AGG$ can be applied, such as mean, pooling, graph convolution or Long Short-Term Memory (LSTM). For our experiments, we selected the mean aggregation for simplicity, as adopted in~\cite{lo2021graphsage}. The mean aggregation function simply computes the average of edge features over the sampled neighborhood of the nodes. 

Then, to combine the aggregated information $\mathbf{h}_{\mathcal{N}_v}^k$ and the node embedding in previous layer $\mathbf{h}_{v}^{k-1}$,  they are concatenated and multiplied with a trainable weight matrix $W_k$. Finally, the node embeddings are yielded after activation $\sigma$: 

\begin{equation}
    \mathbf{h}_{v}^k = \sigma(\mathbf{W}^k\cdot [\mathbf{h}_{v}^{k-1} \| \mathbf{h}_{\mathcal{N}_v}^k]),
\end{equation}

\noindent where $\|$ represents concatenation. Notice that the nodes in network graph are featureless. They are initialized with all-one vectors $\mathbf{h}_v^{0} = \mathbf{1}$ with the dimension equal to the number of edge features. After the aggregation, the final edge embeddings at the $K^{th}$ layer are calculated as the concatenation of node embeddings from the endpoints in the original E-GraphSage, as shown in Equation~\ref{eq:four}. 

\begin{equation}\label{eq:four}
    \mathbf{z}_{uv} = \mathbf{h}_u^K\| \mathbf{h}_v^K, \forall{uv} \in \mathcal{B}
\end{equation}

However, original edge features may be diluted during the mean aggregation and are not well represented in the final edge embeddings. To address this issue, we propose a modification at this stage. To obtain improved final edge embeddings, we propose concatenating the node embeddings from the endpoints and the original edge features as well, as shown in line 9 in Algorithm~\ref{alg:eagg}. Equation~\ref{eq:concat1} shows this new concatenation procedure.

\begin{equation}\label{eq:concat1}
    \mathbf{z}_{uv} = \Vert(\{\mathbf{h}_u^K, \mathbf{h}_v^K, \mathbf{e}_{uv}\}), \forall{uv} \in \mathcal{B}
\end{equation}

This represents the final edge embeddings in our modified E-GraphSAGE algorithm. Notice that this is very similar to a residual learning~\cite{he2016deep}, where $\mathbf{z}_{uv} = \mathcal{F}(\mathbf{e}_{uv}) + \mathbf{e}_{uv}$ with residual function $
\mathcal{F}$, except that the summation is replaced by concatenation. The proposed modification to the E-GraphSAGE algorithm provided a significant improvement in the experiments carried out, showing the clear advantage of using this new solution. 

\begin{algorithm}[ht]
\SetAlgoLined
\KwInput{   Graph $\mathcal{G(V,E)}$\;  
\Indp\Indp edge minibatch $\mathcal{B}$; 2-hop neighborhood $\mathcal{B}^0$\;
node set $V(\mathcal{B})=\{v:v$ is an end point of $uv \in \mathcal{B}\}$\;
input edge features \{$\mathbf{e}_{uv},\forall{uv} \in{\mathcal{B}} $\}\;
input node features $\mathbf{x}_{v}=\mathbf{1}$\;
depth or layer $K$; non-linearity $\sigma$\;
weight matrices $\mathbf{W}^k,\forall{k}\in \{1,...,K\}$\;
differentiable aggregator functions $AGG_k$\;
neighborhood sampling functions $\mathcal{N}_v:v\rightarrow2^\mathcal{V}$\;
}
\KwOutput{Vector representation $\mathbf{z}_{uv}, \forall{uv} \in \mathcal{B}$}
 \BlankLine
 $\mathbf{h}_{uv}^0\leftarrow\mathbf{e}_{uv},\forall{uv} \in \mathcal{B}^0$\;
 $\mathbf{h}_v^0\leftarrow\mathbf{x}_v,\forall{v} \in V(\mathcal{B}^0)$\;
 \For{$k=1,...,K$}{
  \For{$v \in V(\mathcal{B}^k)$}{
  $\mathbf{h}_{v}^k = \sigma(\mathbf{W}^k\cdot [\mathbf{h}_{v}^{k-1} \| \mathbf{h}_{\mathcal{N}_v}^k])$\;
  $\mathbf{h}_{v}^k = \sigma(\mathbf{W}^k\cdot CONCAT(\mathbf{h}_{v}^{k-1}, \mathbf{h}_{\mathcal{N}(v)}^k))$\;
  }
 }
 $\mathbf{z}_{uv} = \Vert(\{\mathbf{h}_u^K, \mathbf{h}_v^K, \mathbf{e}_{uv}\}), \forall{uv} \in \mathcal{B}$\; \tcp{concatenate the original features as well}
 \caption{E-GraphSage minibatch aggregation}
 \label{alg:eagg}
\end{algorithm}

\subsection{The E-ResGAT Algorithm}\label{sec:3_3}


In the E-GraphSage algorithm, all the neighboring edges are treated equally. However, it is more natural to consider different weights for different neighbors. Hence, we decided to apply the graph attention network~\cite{velivckovic2017graph} with residuals to our problem, which additionally learns the weighted aggregation. Another key difference between E-ResGAT and E-GraphSAGE algorithms is that E-ResGAT operates on the line graph $\mathcal{G'(V',E')}$ instead of the original bipartite graph $\mathcal{G(V,E)}$, where $\mathcal{V'=E} $ as specified in Figure~\ref{fig:bigraph}. This way, the model does not split the neighborhood with regard to the endpoint, thus allowing for a neighborhood of approximately twice the size to be sampled which means that more useful information is aggregated.

\begin{figure*}[htb]
    \centering
    \includegraphics[width=\textwidth]{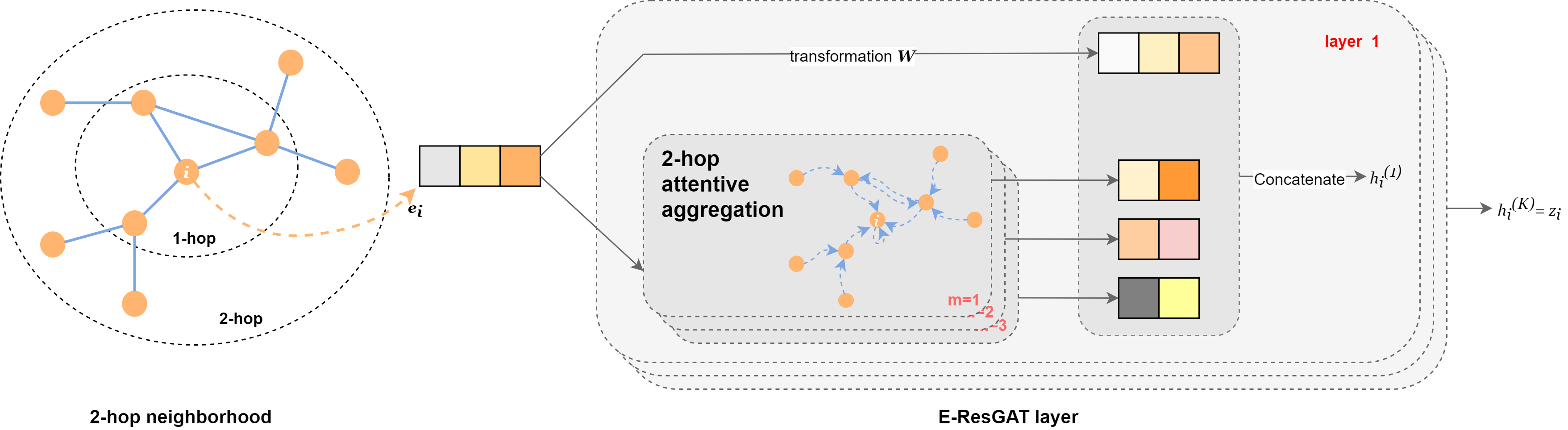}
    \caption{The overall architecture of E-ResGAT.}
    \label{fig:resgat}
\end{figure*}

The algorithm first constructs a 2-hop full neighborhood of a batch. We omit a detailed description of this step due to its similarity to the sampling step described in Section~\ref{sec:egrap}. The only difference is that we use the full neighborhood of nodes in the line graph (Figure~\ref{fig:bigraph} on the right hand side). Then, an E-ResGAT layer aggregates the neighboring information using the attention mechanism. Moreover, inspired by the performance improvements observed with the modification made in E-GraphSAGE, we also concatenated a transformation of the original node feature in each layer. Thanks to such residual learning, the E-ResGAT can go deeper than the vanilla GAT, and the stacked aggregation layers will gradually refine the node embeddings. The overall architecture of the proposed E-ResGAT model is described in Figure~\ref{fig:resgat}, and the mini-batch version of the E-ResGAT aggregation is shown in Algorithm~\ref{alg:egat}.

\paragraph{\textbf{\textup{Aggregation with Residual}}}
We next describe the E-ResGAT aggregation layer in detail. The inputs are node features $\mathbf{h}_v$ in a line graph, where $v \in \mathcal{B}^0$. The layer outputs are the aggregated node embeddings $\mathbf{h'}_v$. 

For each node $v$, a weighted average of the neighboring features is first computed, and then concatenated with the transformed node features $\mathbf{e}_v$. This works for the highly imbalanced data in malware detection, as it can prevent jeopardizing the performance when the neighborhood of a node is mostly occupied by majority class and the node embedding is not well represented. At the $k^{th}$ layer, the attention-based aggregation with residual can be expressed as:
\begin{equation}
    \mathbf{h}^k_v = \sigma(\sum_{u \in \mathcal{N}_v}\alpha_{uv}\mathbf{W}\mathbf{h}_u^{k-1}) \Vert (\mathbf{W'e}_v) ,
\end{equation}

\noindent where $\alpha_{uv}$ is the attention coefficient assigned to the edge $e_{uv}$ in the line graph, and $\mathbf{W}$ is a shared linear transformation across the layer that maps the input features into a lower dimension. The attention coefficient $\alpha_{uv}$ can be learned simply by a feedforward neural network $\mathbf{a[Wh}_u\|\mathbf{Wh}_v]$, where $\mathbf{a}$ is a weight vector. Then, the attention coefficients for all node pairs are obtained after LeakyReLU activation~\cite{maas2013rectifier} and a softmax function, as shown in Equation~\ref{eq:seven}.
\begin{equation}\label{eq:seven}
    \alpha_{uv} = \frac{\textup{exp}\left(\textup{LeakyReLU}(\mathbf{a[Wh}_u\|\mathbf{Wh}_v])\right)}{\sum_{i \in N_v}\textup{exp}\left(\textup{LeakyReLU}(\mathbf{a[Wh}_i\|\mathbf{Wh}_v])\right)}
\end{equation}

Similar to GAT, we can apply multi-head attentions to increase the capacity of the E-ResGAT model. Thus, a multi-head E-ResGAT aggregation is formulated as:
\begin{equation}
  \begin{gathered}\
    \mathbf{h}^k_{\mathcal{N}_v} = \underset{m=1}{\overset{M}{\Vert}}\sigma(\sum_{u \in \mathcal{N}_v}\alpha_{uv}^{m}\mathbf{W}^{m}\mathbf{h}_u^{k-1}) \\
    \mathbf{h}^k_v = \mathbf{h}^k_{\mathcal{N}_v} \Vert \mathbf{W'}\mathbf{e}_v,
  \end{gathered}
\end{equation}
where there are $M$ heads of attention, $\alpha_{uv}^m$ are the $m^{th}$ normalized attention coefficients, and $\mathbf{W}^m$
corresponds to $m^{th}$ weight matrix. Note that we concatenate the original node features $e_v$ at the end with a transformation $\mathbf{W'}$. Such transformation can be the average of all $M$ matrices $\overline{\mathbf{W}^m}$. To reduce the computation, we simply implemented a identity mapping.

\begin{algorithm}[ht]
\SetAlgoLined
\KwInput{   Graph $\mathcal{G'(V',E')}$\;  
\Indp\Indp edge minibatch $\mathcal{B}$; 2-hop neighborhood $\mathcal{B}^0$\;
input node features $\mathbf{e}_{v}$\;
layer $K$; number of heads $M$\;
weight matrices $\mathbf{W}^m,\forall{m}\in \{1,...,M\}$\;
non-linearity $\sigma$\;
neighborhood for node $v$ $\mathcal{N}_v:v\rightarrow2^\mathcal{V}$\;
}
\KwOutput{Vector representation $\mathbf{z}_{v}, \forall{v} \in \mathcal{B}$}
 \BlankLine
 $\mathbf{h}_{v}^0\leftarrow\mathbf{e}_{v},\forall{v} \in \mathcal{B}^0$\;
 \For{$k=1,...,K$}{
    \For{$m=1,...,M$}{
        $e_{uv} = \textup{LeakyReLU}(\mathbf{a}[\mathbf{W}^m\mathbf{h}^{k-1}_u\Vert\mathbf{W}^m\mathbf{h}^{k-1}_v])$\;
        
        $\alpha_{uv} = \textup{softmax}(e_{uv}) = \frac{\textup{exp}(e_{uv})}{\sum_{i \in N_v}\textup{exp}(e_{iv})}$\;
    }
    $\mathbf{h}^k_{\mathcal{N}_v} = \underset{m=1}{\overset{M}{\Vert}}\sigma(\sum_{u \in \mathcal{N}_v}\alpha_{uv}^{m}\mathbf{W}^{m}\mathbf{h}_u^{k-1})$\;
    $\mathbf{h}^k_v = \mathbf{h}^k_{\mathcal{N}_v} \Vert \mathbf{W'}\mathbf{e}_v$\;
 }
 $\mathbf{z}_v = \mathbf{h}^K_{\mathcal{N}_v} \Vert \mathbf{W'}\mathbf{e}_v, \forall{v} \in \mathcal{B}$\;
 
 \caption{E-ResGAT minibatch aggregation}
 \label{alg:egat}
\end{algorithm}



\section{Experimental Evaluation}\label{sec:4}
In order to evaluate the performance of the two proposed models, we carried out our experiments on four intrusion detection datasets. First, a brief description of all the considered datasets is presented. Then, we discuss the detailed experimental setting, including the metrics and implementation. Subsequently, we present the results of both binary and multiclass classification problems, followed by an extensive analysis and discussion on efficiency and learned features. 

\subsection{Data Sets}

For our experiments we selected four intrusion detection datasets. Their main characteristics are described in Table~\ref{tab:ds1}. The datasets chosen represent multiclass problems where the percentage of benign cases is much higher than the percentage of attacks. For dataset CIC-DarkNet\footnote{https://www.unb.ca/cic/datasets/darknet2020.html}~\cite{habibi2020didarknet} we used the entire data available. However, for the remaining datasets we only used a sample of the data available on the webpage. We used the train/test files provided for dataset ToN-IoT\footnote{https://research.unsw.edu.au/projects/toniot-datasets}\cite{alsaedi2020ton_iot}, while for datasets UNSW-NB15\footnote{https://research.unsw.edu.au/projects/unsw-nb15-dataset}\cite{moustafa2015unsw} and CSE-CIC-IDS\footnote{https://www.unb.ca/cic/datasets/ids-2018.html}\cite{sharafaldin2018toward} we use only a sample of the data available. In particular, we selected roughly 25\% and 3\% of the data available for these two datasets, respectively. The motivation for this is related with both the large size of the datasets and the generation of different settings with regards to the percentages of the majority class (normal) cases. 

Table~\ref{tab:classesds} shows the distribution obtained for all classes in each dataset. We observe that these intrusion detection datasets have different characteristics. All are multiclass problems with one majority class (the normal class) and multiple minority classes (the different attacks). The majority class has a different representation in each dataset ranging from 96,83\% to 65.07\%. The minority classes also exhibit different distributions ranging between 13.04\% (DDOS attack on CES-CIC-IDS dataset) and 0.003\% (Worms attack in UNSW-NB15 dataset). This provides a challenging scenario for our experiments.


\begin{table*}[htp]
\centering
\begin{tabular}{@{}llllll@{}}
\toprule
\textbf{Dataset} & \textbf{No. Examples} & \textbf{Normal(\%)} & \textbf{No. Classes} & \textbf{No. Features} & \textbf{Reference} \\ \midrule
UNSW-NB15 & 700001 & 96.83 & 10 & 43 & \cite{moustafa2015unsw} \\
CIC-DarkNet & 141530 & 82.82 & 9 & 77 & \cite{habibi2020didarknet} \\
CSE-CIC-IDS & 250928 & 70.61 & 7 & 77 & \cite{sharafaldin2018toward} \\
ToN-IoT & 461043 & 65.07 & 10 & 39 & \cite{alsaedi2020ton_iot} \\
 \bottomrule
\end{tabular}
\caption{Overview of the datasets used in our experiments.}
\label{tab:ds1}
\end{table*}

\begin{table*}[htp]
\fontsize{9}{11}\selectfont
\centering
\begin{tabular}{@{}l@{}l@{}l@{}l@{}l@{}l@{}l@{}l@{}l@{}l@{}l@{}}
\toprule
\textbf{Dataset}~~ & \multicolumn{10}{c}{\textbf{Classes (names and \%)}} \\ \midrule
\multirow{2}{*}{UNSW-NB15} & Normal ~ & Exploits & Recon. ~ & DoS & Generic ~ & Shellcode ~ & Fuzzers ~& Worms & Backd. ~ & Analysis \\
\multicolumn{1}{c}{} & 96.83 & 0.773 & 0.251 & 0.167 & 1.07 & 0.032 & 0.722 & 0.003 & 0.076 & 0.075 \\
 \cmidrule{2-11}
\multirow{2}{*}{CIC-DarkNet} & Normal & Audio str ~ & Brows. & Chat & File tr. & Email & P2P & Video str & VOIP &  \\
 & 82.82 & 9.39 & 0.19 & 3.21 & 1.84 & 0.41 & 0.16 & 0.95 & 1.03 &  \\
\cmidrule{2-11}
\multirow{2}{*}{CSE-CIC-IDS ~~} & Normal & BruteF. & DoS & DDoS ~ & Web & Bot & Infiltr. &  &  &  \\
 & 70.61 & 7.28 & 1.01 & 13.04 & 0.37 & 5.42 &2.27 &  &  &  \\
 \cmidrule{2-11}
\multirow{2}{*}{ToN-IoT} & Normal & Scanning & Dos & Inject. & DDos & Password & XSS & Ransomw. ~ & Backd. & MITM \\
 & 65.07 & 4.34 & 4.34 & 4.34 & 4.34 & 4.34 & 4.34 & 4.34 & 4.34 & 0.22 \\
 \bottomrule
\end{tabular}
\caption{Distribution of the classes of the selected datasets.}
\label{tab:classesds}
\end{table*}

\subsection{Experimental Setting}

\paragraph{\textbf{\textup{Evaluation Metrics}}}
The overall accuracy performance of the proposed models is measured through the F1-score, which is expressed below:

\begin{equation}
    F1\mbox{-}score = 2\times\frac{Recall\times Precision}{Recall + Precision},
\end{equation}
where $Precision$ measures the ability of an intrusion
detection system to identify only the attacks, while $Recall$ can
be thought as the system’s ability to find all the attacks. The
higher the F1-score, the better the balance between $Precision$ and
$Recall$ achieved by the model. We initially compare the weighted F1-scores as overall model performance for both binary and multi-classification tasks. The weight coefficients used are the class ratios. However, we do not want the high weighted F1-score in cost of the minority accuracy. Thus, a macro F1-score is considered as well. The F1-scores of all classes are averaged evenly in calculation of macro F1-score. This way, we do not discriminate the classes by their percentage and raise the importance of minority classes. Both metrics are computed on test sets of corresponding datasets.

The efficiency performance is evaluated with the time spent on training the proposed models. The time is collected from a Windows machine with an Intel(R) Core(TM) i5-8250U CPU @ 1.60GHz and 8 GB RAM. The batch training time is measured in seconds.

\paragraph{\textbf{\textup{Models and Hyperparameters}}}
For the modified E-GraphSAGE, a 2-layer model, followed by softmax classifier, is used for edge classification. The original E-GraphSAGE is compared with the modified version as a baseline model. For all the E-GraphSAGE models, either modified or not, we used mean function as aggregation, ReLU~\cite{nair2010rectified} as the non-linearity, neighborhood as 2-hop and sample size as 8 for both layers. 

For E-ResGAT, a deeper model with 6-head attention at each layer is used for node classification. The number of layers is selected based on the best performing hyperparameter value. To make a fair comparison, we also implemented a vanilla GAT model as a baseline to illustrate the impact of residual modules. In order to avoid overfitting, dropout is applied to attention coefficients. It is also applied to the input in original GAT model~\cite{velivckovic2017graph}. We did not adopt the dropout step, as the number of features in the considered datasets is already very small. Similar to the original GAT, ELU~\cite{clevert2015fast} is used as the non-linearity. 

All models were implemented in Pytorch~\cite{paszke2019pytorch} using Adam optimizer~\cite{kingma2014adam}. In order to give a fair comparison, these models share identical minibatch iterators, loss function, learning rates and the splitting of train, validation and test sets. In practice, we used minibatchs of size 500, cross-entropy loss and 5/2/3 for train/validation/test ratio. Learning rates ($lr$) were chosen from the best setting for each dataset according to the performance on the validation set. This way, the learning rates used are as follows: $lr=0.007$ for UNSW-NB15 and $lr=0.003$ for both CIC-DarkNet and CSE-CIC-IDS, while $lr=0.01$ for ToN-IoT. All the models were trained for two epochs on the training set.

\begin{table*}[htp]
\centering
\begin{threeparttable}
\fontsize{9}{11}\selectfont
\centering
\begin{tabular}{@{}lllllll@{}}
\toprule
\multirow{2}{*}{\textbf{Dataset}} & \multirow{2}{*}{\textbf{Algorithm}} & \multicolumn{2}{c}{\textbf{F1 Weighted}} &  & \multicolumn{2}{c}{\textbf{F1 Macro}} \\ \cmidrule(lr){3-4} \cmidrule(l){6-7} 
 &  & Binary & Multi &  & Binary & Multi \\ \midrule
\multicolumn{1}{c}{\multirow{4}{*}{UNSW-NB15}} & E-GraphSAGE & 0.9944 & \textbf{0.9868}*&  & \textbf{0.9539} &  \textbf{0.4091}* \\
\multicolumn{1}{c}{} & E-GraphSAGE M & \textbf{0.9950}* & 0.9837 &  & 0.9519 & 0.3631 \\
\multicolumn{1}{c}{} & GAT & \textbf{0.9948} & 0.9855 &  & \textbf{0.9589}* & 0.3830 \\
\multicolumn{1}{c}{} & E-ResGAT & 0.9946 & \textbf{0.9858} &  & 0.9578 & \textbf{0.3865} \\
\midrule
\multirow{4}{*}{CIC-DarkNet} & E-GraphSAGE & 0.8693 & 0.8093 &  & 0.7465 & 0.3554 \\
 & E-GraphSAGE M & \textbf{0.9150} & \textbf{0.8790} & \textbf{} & \textbf{0.8519} & \textbf{0.4248} \\
 & GAT & 0.9206 & 0.8749 &  & 0.8628 & 0.4343 \\
 & E-ResGAT & \textbf{0.9232}* &  \textbf{0.8807}* & \textbf{} &  \textbf{0.8666}* &  \textbf{0.5087}* \\
\midrule
\multirow{4}{*}{CSE-CIC-IDS} & E-GraphSAGE & 0.9082 & 0.8774 &  & 0.8657 & 0.5505 \\
 & E-GraphSAGE M & \textbf{0.9632} & \textbf{0.9557}* &  & \textbf{0.9554} & \textbf{0.7591} \\
 & GAT & 0.9634 & 0.9494 &  & 0.9554 & 0.7688 \\
 & E-ResGAT &  \textbf{0.9650}* & \textbf{0.9531} & \textbf{} &  \textbf{0.9573}* & \textbf{0.9101}* \\ \midrule
\multirow{4}{*}{ToN-IoT} & E-GraphSAGE & 0.9953 & 0.9384 &  & 0.9949 & 0.7695 \\
 & E-GraphSAGE M &  \textbf{0.9988}* & \textbf{0.9851} & \textbf{} & \textbf{0.9987}* & \textbf{0.9385} \\
 & GAT & 0.9976 & 0.9954 &  & 0.9973 & 0.9776 \\
 & E-ResGAT &  \textbf{0.9988}* &  \textbf{0.9970}* & \textbf{} & \textbf{0.9986} &  \textbf{0.9930}* \\  \bottomrule
\end{tabular}
\caption{Weighted and Macro F1 measure results on the four selected datasets. Numbers in bold are the higher scores achieved in each pair of algorithms, while those with the asterisk(*) refer to the highest scores among all algorithms.}
\label{tab:res*}
\end{threeparttable}
\end{table*}

\begin{table*}[htp]
\fontsize{9}{11}\selectfont
\resizebox{\textwidth}{!}{%
\begin{tabular}{@{}llllllllllll@{}}
\toprule
\multirow{2}{*}{\textbf{Dataset}} & \multirow{2}{*}{\textbf{Algorithm}} & \multicolumn{10}{c}{\textbf{Per class F1-score}} \\ \cmidrule(l){3-12} 
 &  & 0 & 1 & 2 & 3 & 4 & 5 & 6 & 7 & 8 & 9 \\ \midrule
\multirow{4}{*}{UNSW-NB15} & E-GraphSAGE & 0.9970 & 0.6386 & 0.7904 & 0.0515 & \textbf{0.9472} & \textbf{0.0900} & \textbf{0.5763} & 0.0000 & 0.0000 & 0.0000 \\
 & E-GraphSAGE M & 0.9970 & 0.5832 & \textbf{0.8383} & 0.0316 & 0.9227 & 0.0000 & 0.2578 & 0.0000 & 0.0000 & 0.0000 \\
 & GAT & 0.9970 & 0.6280 & 0.4280 & \textbf{0.0675} & 0.9378 & 0.0000 & 0.5221 & 0.0000 & \textbf{0.2500} & 0.0000 \\
     & E-ResGAT & \textbf{0.9973} & \textbf{0.6540} & 0.4886 & 0.0490 & 0.9252 & 0.0000 & 0.5242 & 0.0000 & 0.2268 & 0.0000 \\
      \midrule
\multirow{4}{*}{CIC-DarkNet} & E-GraphSAGE & 0.9355 & 0.1575 & 0.4426 & 0.3351 & 0.2087 & 0.1936 & \textbf{0.6971} & \textbf{0.3090} & 0.0000 &  \\
 & E-GraphSAGE M & 0.9450 & \textbf{0.6684} & 0.3158 & \textbf{0.7424} & 0.3093 & 0.0794 & 0.5328 & 0.1809 & 0.0492 &  \\
 & GAT & 0.9484 & 0.6453 & 0.4179 & 0.4960 & 0.2650 & 0.2960 & 0.6210 & 0.0770 & 0.1420 &  \\
 & E-ResGAT & \textbf{0.9511} & 0.6209 & \textbf{0.4978} & 0.6079 & \textbf{0.3672} & \textbf{0.3345} & 0.6784 & \textbf{0.3090} & \textbf{0.2113} &  \\
 \midrule
\multirow{4}{*}{CSE-CIC-IDS} & E-GraphSAGE & 0.9329 & 0.9956 & 0.0000 & 0.4712 & 0.0000 & 0.9030 & 0.0000 &  &  &  \\
 & E-GraphSAGE M & 0.9752 & \textbf{0.9964} & 0.9010 & 0.9839 & 0.5817 & \textbf{0.9925} & \textbf{0.0549} &  &  &  \\
 & GAT & 0.9740 & 0.9789 & 0.9286 & 0.9895 & 0.5934 & 0.9175 & 0.0000 &  &  &  \\
 & E-ResGAT &\textbf{ 0.9760} & 0.9870 & \textbf{0.9561} & \textbf{0.9909} & \textbf{0.6108} & 0.9400 & 0.0000 &  &  &  \\
 \midrule
\multirow{4}{*}{ToN-IoT} & E-GraphSAGE & 0.9931 & 0.7297 & 0.9196 & 0.5503 & 0.0000 & 0.9955 & 0.7225 & 0.9625 & 0.9616 & 0.8606 \\
 & E-GraphSAGE M & 0.9949 & 0.9103 & 0.9645 & 0.9072 & 0.7028 & \textbf{0.9992} & 0.9557 & 0.9844 & 0.9887 & 0.9771 \\
 & GAT & 0.9947 & 0.9865 & 0.9810 & 0.9884 & 0.8532 & 0.9985 & 0.9898 & \textbf{0.9967} & 0.9928 & 0.9946 \\
 & E-ResGAT & \textbf{0.9951} & \textbf{0.9907} & \textbf{0.9897} & \textbf{0.9937} & \textbf{0.9793} & 0.9987 & \textbf{0.9946} & 0.9964 & \textbf{0.9956} & \textbf{0.9963}\\
\bottomrule
\end{tabular}
}
\caption{F1-score results per class on the four selected datasets. Numbers in bold refer to the highest scores achieved in each class among all the implemented algorithms.}
\label{tab:res2}
\end{table*}

\subsection{Results and Discussion}\label{sec:4_3}
\paragraph{\textbf{\textup{Classification}}}
We first carry out a binary classification evaluation, assessing whether a flow belongs to a normal or malicious class and then move to a multiclass scenario where we identify both the benign and the individual attack classes. As mentioned above, we report F1 weighted and F1 macro scores for binary classification and the muticlass scenario. Table~\ref{tab:res*} displays the F1 weighted and F1 macro results obtained with the four algorithms in the four datasets
In order to fairly assess the benefits of the residual features, comparison is made within each pair of modified and original models. As observed from Table~\ref{tab:res*}, the two models we propose clearly outperform the original ones in three out of four datasets, in terms of both F1 weighted and F1 macro scores. 
The results are different for dataset UNSW-NB15 where the proposed alternatives do not win every time. Still, in this case, we observe that E-ResGAT yields better scores than GAT in the multiclass classification setting for both metrics. Moreover, on this dataset, for the binary F1 Macro scenario, the differences observed between the results of our models and the competitor models are very small (0.2\% on GraphSAGE-based models and 0.11\% for GAT-based models). We must also highlight that a similar behaviour was observed for the UNSW-NB15 in the original E-graphSAGE experiments~\cite{lo2021graphsage}, when it was compared against other non-graph based start-of-the-art models. Hence, we hypothesize that graph-based models may not be the most suitable solution for this particular highly imbalanced dataset. Finally, it is also noticeable the clear positive impact of embedding an attention mechanism in the algorithms. In effect, if we compare between the two groups of models, GAT-based and GraphSAGE-based, it becomes evident that GAT-related models generally beat GraphSAGE-related models, indicating the power of the attention mechanism. 

To provide a more detail overview of the results achieved when tackling these problems as a multiclass task, we provide in Table~\ref{tab:res2} the individual F1-score for each class obtained by the four models tested in all the selected datasets. The proposed E-ResGAT gives the best per-class F1 scores in most of the cases. To be more specific, E-ResGAT outperforms all the other models on 20 out of 34 classes in total. Class 7 and 9 in UNSW-NB15 is excluded as none of the models can distinguish them given such a low ratio in the dataset. For the classes where this is not the case, such as class 1 and 3 in CIC-DarkNet and class 1, 5 and 6 in CSE-CIC-IDS, etc, the proposed modified E-GraphSAGE achieves the best scores. So, our results successfully demonstrate the beneficial effect of residual features in most of our considered datasets, as expected in Section~\ref{sec:3_3}. Furthermore, it is worth noting the improvements achieved on individual classes: there is at least 1\% increment from E-ResGAT w.r.t. other models in most classes. Moreover, our proposed models display an overall impressive boost in the performance of the classes in which the original E-GraphSAGE shows near zero F1-scores, once again reflecting the significance of residual features.

\paragraph{\textbf{\textup{Efficiency}}}
The efficiency of the implemented models is evaluated by the time spent in terms of seconds in training a single batch of 500 network flow samples. The overall average training time of the two GraphSage-based models and the two GAT-based models is summarised in Table~\ref{tab:eff}. The efficiency within each type of models (the GraphSAGE-based and the GAT-based) does not differ very much. However, we observe that the total time of the GAT-based models we implemented is over 50 times that of GraphSAGE-based models. We must highlight that the actual training time of GAT-based models is only about 3~5 times more than that of GraphSAGE-based models as shown by the right most value in Table~\ref{tab:eff}. The main reason that justifies this is the transformation from the original bipartite graph structure to its line graph structure, which is required for GAT-based models as discussed in Section~\ref{sec:3_1}. Within this transformation, we select all the edges that share common endpoints with a batch of edges to construct the full neighborhood of that batch, as opposed to the neighborhood sampling procedure applied in E-GraphSAGE. The time needed for the transformation is in $\mathcal{O(|B|} \overline{D}^2)$, where $\mathcal{|B|}$ is the batch size and $\overline{D}$ is the averaged degree. If available memory permits, one can store the line graph in disk and significantly save the training time. The optimization of the graph structure transformation is left for future work.

\begin{table}[htp]
\centering
\begin{tabular}{@{}cccc@{}}
\toprule
\textbf{Dataset} & \textbf{GraphSAGE-based} & \textbf{GAT-based} \\ \midrule
UNSW-NB15 & 0.171 & 5.65 (5.05 + 0.602) \\ 
CIC-DarkNet & 0.127 & 4.48 (4.12 + 0.361)\\ 
CSE-CIC-IDS & 0.119 & 5.72 (5.33 + 0.656) \\ 
ToN-IoT & 0.154 & 7.17 (6.52 + 0.651) \\ 
 \bottomrule
\end{tabular}
\caption{Averaged time spent (in seconds) in training a batch of size 500 using GraphSAGE-based and GAT-based algorithms. The total time of GAT models is further divided into time for line graph transformation (left in brackets) and actual training time (right in brackets).}
\label{tab:eff}
\end{table}



\begin{figure}[htb]
    \centering
    \includegraphics[width=0.48\textwidth]{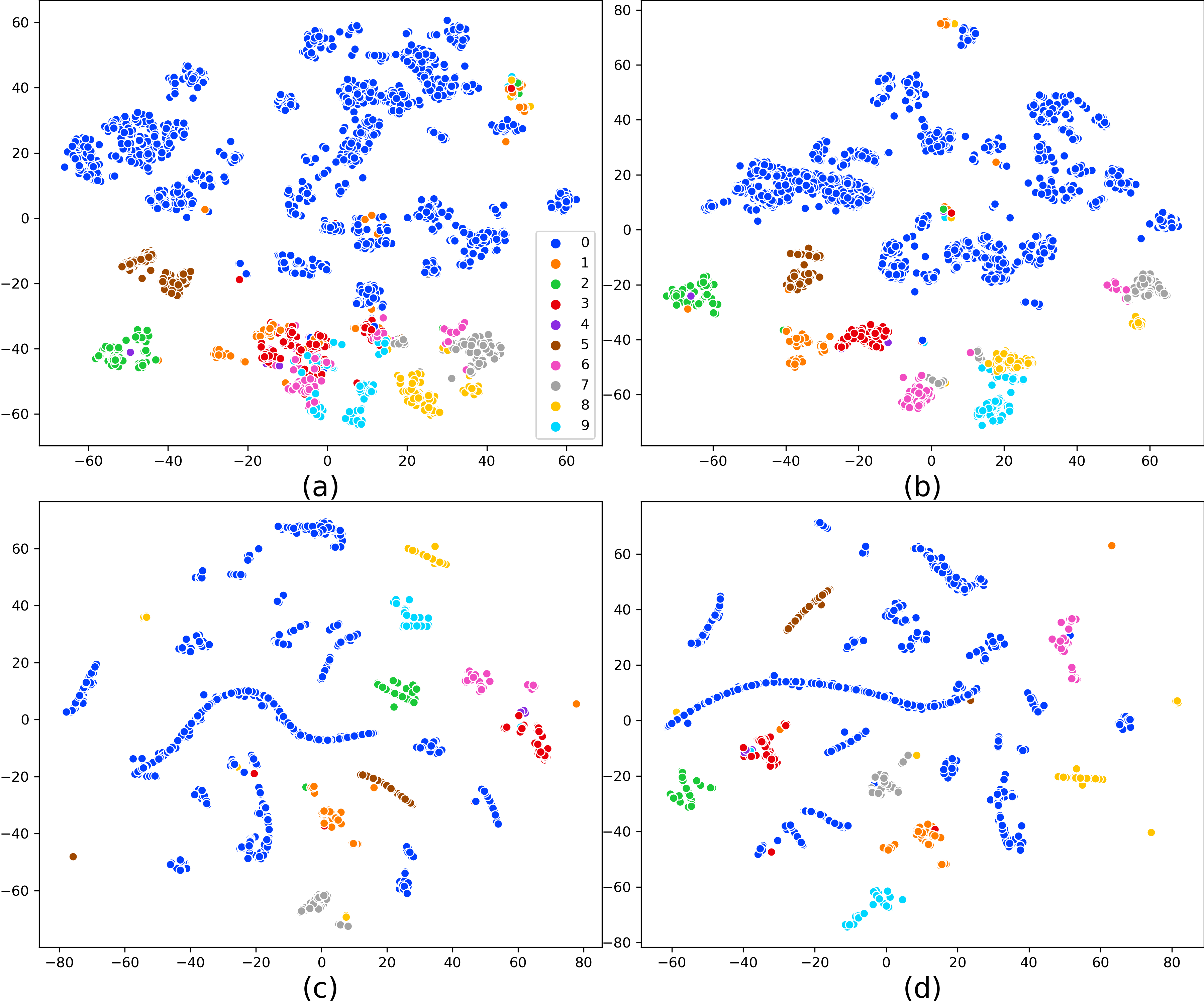}
    \caption{A t-SNE visualization of learned feature representations of all four models' last hidden layer on the ToN-IoT dataset. (a) E-GraphSAGE (b) E-GraphSAGE Modified (c) GAT (d) E-ResGAT}
    \label{fig:tsne}
\end{figure}

\paragraph{\textbf{\textup{Learned Feature Representation}}}
Finally, we provide a visualization of the learned feature representations from the last layer of all four models using the t-SNE method~\cite{van2008visualizing}. The results of dataset ToN-IoT are shown in Figure~\ref{fig:tsne}. The representation exhibits discernible clustering in the projected two dimensional plane. It can be observed that each pair of models learns similar representations on ToN-IoT. Moreover, an interesting pattern is detected by the GAT-related models ((c) and (d) in Figure~\ref{fig:tsne}). Instead of globular shaped clusters, the t-SNE plots of GAT-related models yield long curved clusters. We hypothesize that this behaviour may be attributed to the sequential patterns learned by the attention mechanism present in both algorithms. Comparing between two pairs of models, we can see that there exists a higher class separability in the models with residuals ((b) and (d) in Figure~\ref{fig:tsne}), which also confirms the model’s discriminatory ability across the ten different classes of ToN-IoT dataset. 

\begin{figure}[htb]
    \centering
    \includegraphics[width=0.48\textwidth]{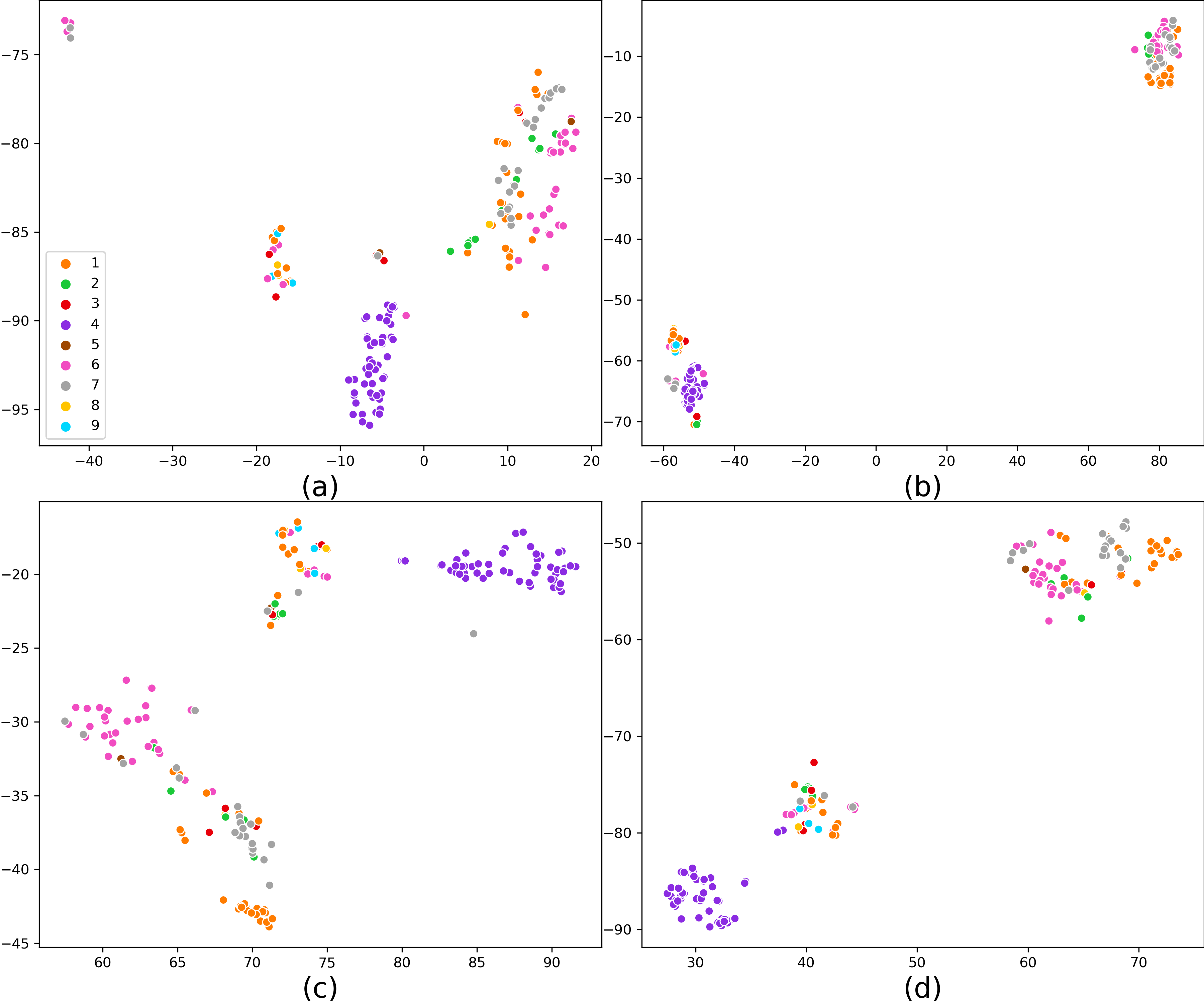}
    \caption{A t-SNE visualization of learned feature representations on the UNSW-NB15 dataset. Class 0 is omitted for clarity. (a) E-GraphSAGE (b) E-GraphSAGE Modified (c) GAT (d) E-ResGAT}
    \label{fig:tsne2}
\end{figure}

Similar results can be found in the remaining datasets, with the exception of UNSW-NB15. Figure~\ref{fig:tsne2} displays the t-SNE results of UNSW-NB15. Due to the high class imbalance, the normal network flow (class 0) is deliberately omitted from the figures to allow a clearer visualization of the minority classes. The key difference of the results lies in the plots of GAT-related models. In this dataset we no longer observe the long curved clusters shown in the other datasets. We hypothesize that the reason for this difference is associated to the inability of the attention mechanism to learn sequential patterns given a highly imbalanced neighborhood with around 95\% normal network flows. Furthermore, there is no significant difference in the class separability among all the four models. While these models perform well in predicting class 4 (Generic attacks), they are less effective to distinguish the other minority classes, which is in line with the experimental results previously discussed.

\section{Conclusions and Future Work}\label{sec:5}
In this paper, we propose two graph-based solutions for intrusion detection. The first model extends the idea of E-GraphSAGE~\cite{lo2021graphsage}, by adding residual connections to the output layer of the E-GraphSAGE. The proposed modified version of E-GrapgSAGE achieves better results when compared against the original version. Inspired by the improved performance, we present edge-based residual graph attention networks (E-ResGATs) in an attempt to solve intrusion detection tasks and deal with class imbalance issues. E-ResGAT introduces residual connections and attention mechanisms to E-GraphSAGE. This results in a more robust graph-based intrusion detection system that assigns larger weights to similar network flows, or at least remain unperturbed under an extremely imbalanced neighborhood. The improved performance in the classification of minority classes validates this point. An extensive set of experiments on four intrusion detection datasets demonstrates the robustness of E-ResGAT in most cases. Our results reveal the potential of GNNs under intrusion detection tasks and the promising future research in this field. We believe that GNNs, with careful modifications, can be widely applied to other areas or different tasks in cybersecurity.

Regarding future explorations, first, we plan to optimize the construction of batch neighborhood to speed up the proposed E-ResGAT model. Another interesting direction is to extend the model to effectively tackle even more imbalanced datasets in intrusion detection, possibly combining it with synthetic data generation techniques (e.g.~\cite{lee2021gan}). 


\section*{Code Availability}

We release a Python implementation of all the four algorithms discussed in this paper for reproducing our work in \href{https://anonymous.4open.science/r/E-ResGAT-552C/README.md}{the anonymous Github repository}. The repository contains the four considered datasets, all the models, a README file with instructions and the code for t-SNE visualization.

\bibliographystyle{plain}
\bibliography{mybibliography}

\end{document}